\documentclass[prb,aps,epsf,twocolumn,floatfix]{revtex4-1}
\usepackage{graphicx}
\usepackage{epsfig}
\usepackage{latexsym}
\usepackage{amsmath,amsfonts,amssymb}
\usepackage{color}
\usepackage{array}

\newcommand{\e}{\varepsilon}

\begin{document}
\title{Erratum: Engineering a p+ip Superconductor: Comparison of Topological Insulator and Rashba Spin-Orbit Coupled Materials [Phys. Rev. B 83, 184520 (2011)]}
\author{Andrew C. Potter and Patrick A. Lee}
\affiliation{ Department of Physics, Massachusetts Institute of
Technology, Cambridge, Massachusetts 02139}

\maketitle

In Ref. \onlinecite{Potter11} (henceforth referred to as {\bf I}), we analyzed the effects of disorder on proposals to create an effective $p+ip$ superconductor from a magnetized two-dimensional electron gas with Rashba spin--orbit coupling (SOC) by placing it on the surface of an ordinary bulk superconductor.  This problem was previously treated numerically in Refs. \onlinecite{PotterMultiband} and \onlinecite{LutchynMultiband}.  In {\bf I}, we pointed out that, since time-reversal symmetry is broken in the surface layer, disorder is generically pair-breaking and tends to suppress the induced superconductivity (SC).    For this system there are three distinct types of disorder: 1) impurities which reside in the SOC surface state, 2) interface disorder and 3) impurities in the bulk superconductor.  Recently the problem of bulk-impurities was addressed again in Ref. \onlinecite{Stanescu11}.  They agree with our general finding that impurity types 1) and 2) are pair breaking but conclude that type 3) is not.  This led us to re--examine our analysis, and we now conclude that our finding concerning type 3) impurities in {\bf I} is incorrect.  As will be explained below, the pair-breaking rate due to bulk impurities is actually negligible.  This removes one barrier to the realization of proximity induced SC in semiconducting nanowires, but our previous conclusion that disorder in the wire and in the interface is detrimental remains the same.

\begin{figure}[tttttt]
\vspace{.1in}
\begin{center}
\includegraphics[width=2.5in]{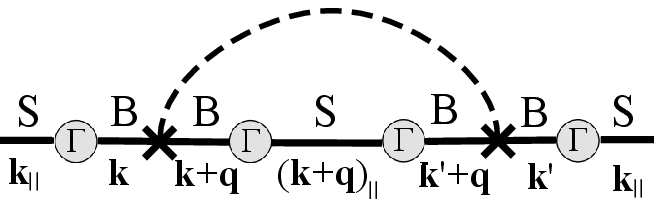}
\end{center}
\caption{Diagrammatic depiction of the pair-breaking process due to bulk impurities. The geometrical constraints on scattering due to the different dimensionality of the surface and bulk (see Fig. \ref{fig:FermiSurfaceGeometry}) suppress these processes by a factor of $\gamma/\e_F\ll 1$.  Circles with $\Gamma$ show surface--bulk tunneling (S and B label surface or bulk Green's functions), bulk impurities are denoted by $\times$, and the dashed line indicates that both $\times$ refer to the same impurity.}
\label{fig:PairBreakingScatteringProcess}
\end{figure}

The pair-breaking effects of bulk-impurities come from processes like that in Fig. \ref{fig:PairBreakingScatteringProcess}. In this process, an electron tunnels from the surface into the bulk, where it is scattered by a bulk-impurity before returning to the surface.  In our analysis in Appendix A of {\bf I}, we assumed the same Fermi-surface geometry for the surface layer and bulk superconductor.  More realistically, the surface-layer is two-dimensional whereas the bulk superconductor is three-dimensional.  Here we show this mixed dimensionality strongly constrains the available phase-space for these scattering processes, and that the pair-breaking effects of bulk-impurities is negligible.  While we only consider the case of a 2D electron gas with SOC, the following argument can also be applied to a 1D (or quasi-1D) wire in contact with a 3D bulk superconductor.

\begin{figure}[tttttt]
\begin{center}
\includegraphics[width=1.5in]{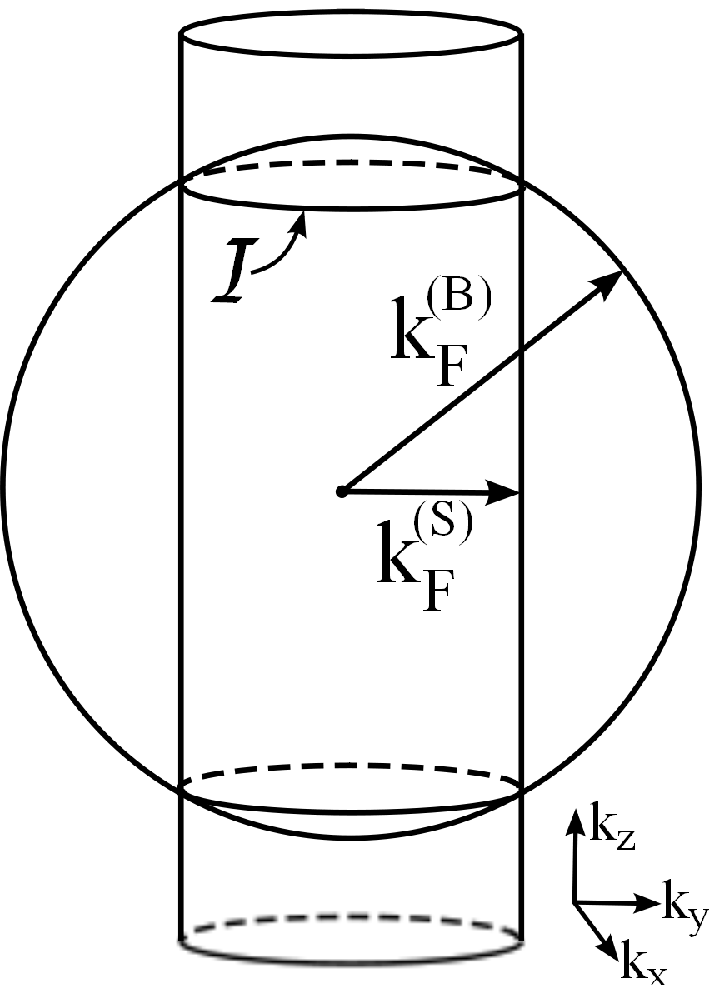}
\end{center}
\vspace{-0.2in}
\caption{Momentum space geometry for surface--bulk tunneling.  The 2D surface Fermi-surface (FS) is extended into a cylinder since tunneling does not conserve the momentum perpendicular to the interface (in the $z$-direction).  Surface--bulk tunneling events involve only states near the intersection, $\mathcal{I}$, of the surface and bulk FS's. }
\label{fig:FermiSurfaceGeometry}
\vspace{-0.2in}
\end{figure}

For a clean interface, the components of momentum parallel  to the surface--bulk interface ($x$ and $y$ components) are conserved whereas the perpendicular ($z$) component is not.
 An electron initially in the surface-layer with momentum $\mathbf{k}_\parallel$ can tunnel into any bulk states with momentum $\mathbf{k} =(\mathbf{k}_\parallel,k_z)$, but pays a large energy cost unless $k_z$ is within $\sim \gamma/v_F$ of the bulk Ferm-surface (FS).  Here $v_F$ is the bulk Fermi-velocity and $\gamma = \pi N_B|\Gamma|^2$ where $N_B$ is the bulk tunneling density of states and $\Gamma$ is the surface--bulk tunneling amplitude.  Once in the bulk the electron can scatter to any momentum $\mathbf{k}+\mathbf{q}$ within $\sim 1/\tau v_F$ of the bulk FS, where $\tau^{-1}$ is the bulk disorder scattering rate.  However, in order to subsequently return to the surface-layer, the in-plane component of $\mathbf{k}+\mathbf{q}$ must again be within $\gamma/v_F$ of the surface FS.  Therefore, the available phase-space for such scattering is $\approx (2\pi k_F)(\frac{1}{\tau v_F})(\frac{\gamma}{v_F})$.  In contrast, the phase space available for arbitrary bulk impurity scattering is $\approx 4\pi k_F^2(\frac{1}{\tau v_F})$.  The pair-breaking scattering rate $\tau_{\text{pb}}^{-1}$ is smaller than the bulk impurity scattering rate $\tau^{-1}$ by the ratio of these two phase-space volumes:
$\tau_{\text{pb}}^{-1}/\tau^{-1} \approx \frac{\gamma}{2v_Fk_F}\sim \frac{\gamma}{\e_F} \ll 1$
where $\e_F$ is the bulk-Fermi energy.  In a typical superconducting metal, $\e_F$ will greatly exceed $\gamma$, hence the pair breaking due to bulk disorder can be safely neglected.  

\begin{figure}[tt]
\vspace{.1in}
\begin{center}
\includegraphics[width=2in]{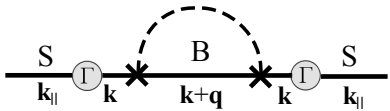}
\end{center}
\caption{Diagrammatic depiction of a non-pair breaking scattering process for surface-electrons due to bulk impurities. Unlike the pair-breaking process shown in Fig. \ref{fig:PairBreakingScatteringProcess}, this process has an unconstrained phase space.}
\label{fig:BulkScatteringProcess}
\end{figure}

Before concluding, we would like to emphasize the distinction between the scattering rate, $\tau_{sb}^{-1}$, for surface-electrons from bulk-impurities and the pair-breaking rate $\tau_{pb}^{-1}$.  The scattering rate, $\tau_{sb}^{-1}$ includes all possible bulk-disorder processes, and is dominated by processes like the one shown in Fig. \ref{fig:BulkScatteringProcess}, where an electron tunnels from surface to bulk, scatters from a bulk impurity and then continues to propagate in the bulk.  This type of scattering is not pair breaking since, after scattering, the electron propagates only in the bulk where time-reversal symmetry is intact and pairing is not disrupted.  Such processes do not suffer the same phase-space restrictions described above, and consequently $\tau_{sb}^{-1}$ can be quite large even though the pair-breaking rate $\tau_{pb}^{-1}$ is small.  Therefore, it is not that the surface-electrons are largely unaffected by bulk impurities, but rather that scattering from these impurities is predominantly non-pair breaking.

\emph{Acknowledgements -- } We thank Roman Lutchyn and Jason Alicea for helpful discussions.



\begin{references}
\bibitem{Potter11}
A.C. Potter and P.A. Lee, Phys. Rev. B {\bf 83}, 184520 (2011)
\bibitem{PotterMultiband}
A.C. Potter and P.A. Lee, Phys. Rev. Lett. 105, 227003 (2010);
\bibitem{LutchynMultiband}
R.M. Lutchyn, T.D. Stanescu, S. Das Sarma, Phys. Rev. Lett. 106, 127001 (2011);
\bibitem{Stanescu11}
T. Stanescu, R. M. Lutchyn, S. Das Sarma arXiv:1106.3078 (2011)
\end{references}
\end{document}